\documentclass[fleqn,twoside]{article}
\usepackage{espcrc2}
\usepackage{graphicx}
\usepackage[figuresright]{rotating}

\newcommand{\AmS}{{\protect\the\textfont2
  A\kern-.1667em\lower.5ex\hbox{M}\kern-.125emS}}

\title{The physics of anomalous glue}
\author{Steven D. Bass\address[ECT]{ECT*, Strada delle Tabarelle 286, I-38050
        Villazzano, Trento, Italy}}

\begin{document}

\begin{abstract}
We give a brief overview of the physics of gluonic degrees of freedom 
associated with the strong (QCD) axial anomaly.
The role of this anomalous glue in the spin structure of the proton and 
$\eta'$-hadron interactions is discussed.
\vspace{1pc}
\end{abstract}

\maketitle

The spin structure of the proton and $\eta'$ physics provide complementary 
windows on the role of gluons in dynamical chiral symmetry breaking.  
The small value of the flavour-singlet axial charge 
$g_A^{(0)}$ extracted from polarized deep inelastic scattering
%the first moment of $g_1$ 
%(the nucleon's first spin dependent structure function) 
\cite{windmolders}
\begin{equation}
\left. g^{(0)}_A \right|_{\rm pDIS} = 0.2 - 0.35
\end{equation}
and the large mass of the $\eta'$ meson point to large violations of OZI 
in the flavour-singlet $J^P=1^+$ channel \cite{bass99,okubo}.
The strong (QCD) axial anomaly is central to theoretical 
explanations of the small value of $g_A^{(0)}|_{\rm pDIS}$ 
(about 50\% of the OZI value 0.6) and the large $\eta'$ mass.
Non-perturbative glue associated with the anomaly may also play a key 
role in the low-energy $\eta'$-nucleon interaction
% which is presently 
%under vigorous experimental investigation at COSY and Jefferson Laboratory.
%It may also play a key role in 
and the QCD structure of 
light-mass exotic states 
with $J^{PC}=1^{-+}$ like those recently observed at BNL and at CERN. 
Here we briefly outline the main details how this anomalous glue may
be central to these phenomena.
We refer to \cite{giusti,uppsala} for more recent detailed 
discussion of the $U_A(1)$ problem and $\eta'$-hadron interactions.

First, let us start with the axial anomaly.
The flavour-singlet axial-vector current 
\begin{equation}
J^{GI}_{\mu5} = \biggl[ \bar{u}\gamma_\mu\gamma_5u
                  + \bar{d}\gamma_\mu\gamma_5d
                  + \bar{s}\gamma_\mu\gamma_5s \biggr]^{GI}_{\mu^2}
\end{equation} 
satisfies the anomalous divergence equation \cite{adler}
\begin{equation}
\partial^\mu J^{GI}_{\mu5}
= 
\sum_{i=1}^{f} 2im_i \bar{q}_i\gamma_5 q_i
+ 3 {\alpha_s \over 8 \pi} G_{\mu \nu} {\tilde G}^{\mu \nu}
\end{equation}
Spontaneous chiral symmetry breaking suggests an octet of 
Goldstone bosons associated with chiral $SU(3)_L \otimes SU(3)_R$ 
plus a singlet boson associated 
with axial U(1) --- each with mass $m^2_{\rm Goldstone} \sim m_q$.
Through coupling to non-perturbative gluon topology via the anomaly
in $J_{\mu 5}^{GI}$ the $\eta$ and $\eta'$ mesons (each with a 
flavour-singlet component) 
acquire additional (larger) 
mass related by 
the Witten-Veneziano formula \cite{giusti}
\begin{equation}
m_{\eta'}^2 + m_{\eta}^2 - 2 m_K^2 
= 
- {6 \over F_{\pi}^2} \chi (0)|_{\rm YM}
\end{equation}
where
\begin{equation}
\chi (0) = \int d^4z \ i \ 
\langle {\rm vac} | T Q(z) Q(0) | {\rm vac} \rangle 
\end{equation}
is the topological susceptibility and
$Q = {\alpha_s \over 4 \pi} G {\tilde G}$ 
is the topological charge density.
The gluonic contribution 
to the $\eta$ and $\eta'$ masses is about 300-400 MeV \cite{uppsala}.

The axial anomaly is important to the spin structure of the proton 
since $g_A^{(0)}$ is measured by the proton forward matrix element 
of $J_{\mu 5}^{GI}$. 
In the QCD parton model the proton's flavour-singlet axial charge 
$g_A^{(0)}$ receives contributions from quark and gluon partons 
\cite{ar} and a possible non-perturbative contribution at $x=0$ 
from gluon topology \cite{bass99}:
\begin{equation}
g_A^{(0)} 
= 
\Biggl(
\sum_q \Delta q - 3 {\alpha_s \over 2 \pi} \Delta g \Biggr)_{\rm partons}
+ \ {\cal C}
\end{equation}
Here ${1 \over 2} \Delta q$ and $\Delta g$ are the amount of spin carried 
by quark and gluon partons in the polarized proton and ${\cal C}$ is the 
topological contribution.
The polarized gluon and possible topology contributions are induced by 
the anomaly.
A positive value of $\Delta g$ acts to reduce the value of 
$g_A^{(0)}|_{\rm pDIS}$ extracted from polarized deep inelastic scattering.
Measuring the gluon polarization $\Delta g$ is the main goal of high-energy
investigations of the spin structure of the proton \cite{gaby}.
A recent QCD motivated fit \cite{jechiel}
to the world $g_1$ data suggests a value 
$\Delta g = 0.63^{+0.20}_{-0.19}$ at 1GeV$^2$, 
in agreement with the prediction \cite{bbs1} 
based on colour coherence and perturbative QCD.
The topological contribution ${\cal C}$ in (6) has support only at 
Bjorken $x$ equal to zero. 
Polarized deep inelastic scattering experiments cannot access $x=0$; 
they measure 
$g_A^{(0)}|_{\rm pDIS} = g_A^{(0)} - {\cal C}$. 
The full $g_A^{(0)}$ may be measured through elastic $Z^0$ 
exchange in $\nu p$ elastic scattering.
A finite value of ${\cal C}$ is natural \cite{bass99} with 
spontaneous $U_A(1)$ symmetry breaking by instantons 
\cite{rjc}
where any instanton induced suppression of 
$g_A^{(0)}|_{\rm pDIS}$ 
(the axial charge carried by partons with finite momentum fraction
 $x>0$)
is compensated by a shift of axial charge to the zero-mode so that
the total axial-charge $g_A^{(0)}$ including ${\cal C}$ is conserved.
In contrast, this zero-mode contribution is not generated 
by instantons
if instantons
explicitly \cite{thooftrep} 
rather than spontaneously break $U_A(1)$ symmetry.
A quality measurement of $\nu p$ elastic scattering would 
provide very valuable information about the nature of 
$U_A(1)$ symmetry breaking in QCD.
If some fraction of the spin of the constituent quark is 
carried by gluon topology in QCD (at $x=0$),  
then the constituent quark model predictions for $g_A^{(0)}$ 
(about 0.6) are not necessarily in contradiction with the 
small value of $g_A^{(0)}|_{\rm pDIS}$ extracted from deep inelastic 
scattering experiments.

The interplay between the proton spin problem and the U(1) problem is 
further manifest in the flavour-singlet Goldberger-Treiman relation 
\cite{venez} which connects $g_A^{(0)}$ with the $\eta'$--nucleon 
coupling constant $g_{\eta' NN}$.
Working in the chiral limit it reads
\begin{equation}
M g_A^{(0)} = \sqrt{3 \over 2} F_0 \biggl( g_{\eta' NN} - g_{QNN} \biggr) 
\end{equation}
where $g_{\eta' NN}$ is the $\eta'$--nucleon coupling constant and 
$g_{QNN}$ is an OZI violating coupling which measures the one 
particle irreducible coupling of the topological charge density 
%$Q = {\alpha_s \over 4 \pi} G {\tilde G}$ 
to the nucleon.
($M$ is the nucleon mass and $F_0 \sim 0.1$GeV renormalises the 
flavour-singlet decay constant.) 
It is important to look for other observables which are sensitive to 
$g_{QNN}$.

Working with the $U_A(1)$--extended chiral Lagrangian for low-energy QCD 
\cite{vecca} one finds a gluon-induced contact interaction in the 
$pp \rightarrow pp \eta'$ reaction close to threshold 
\cite{sb99,uppsala}:
\begin{equation}
{\cal L}_{\rm contact} =
         - {i \over F_0^2} \ g_{QNN} \ {\tilde m}_{\eta_0}^2 \
           {\cal C} \
           \eta_0 \ 
           \biggl( {\bar p} \gamma_5 p \biggr)  \  \biggl( {\bar p} p \biggr)
\end{equation}
Here ${\tilde m}_{\eta_0}$ is the gluonic contribution to the mass of 
the singlet 0$^-$ boson and ${\cal C}$ is a second OZI violating coupling 
which also features in $\eta'N$ scattering.
The physical interpretation of the contact term (8) 
is a ``short distance'' ($\sim 0.2$fm) interaction 
where glue is excited in the interaction region of
the proton-proton collision and 
then evolves to become an $\eta'$ in the final state.
This gluonic contribution to the cross-section 
for $pp \rightarrow pp \eta'$ 
is extra to the contributions associated with meson exchange models.
There is no reason, a priori, to expect it to be small.
The strength of this interaction is presently 
under vigorous 
experimental study at COSY \cite{cosy,cosyprop}. 
Further precision measurements of the low-energy 
$\eta'$-nucleon interaction in 
$\eta'$ photoproduction from nucleons
will soon be available from Jefferson Laboratory \cite{ritchie}.

Recent experiments at BNL \cite{exoticb} and CERN \cite{exoticc} 
have revealed evidence for meson states with exotic quantum 
numbers $J^{PC}=1^{-+}$.
These mesons are particularly interesting because the quantum 
numbers $J^{PC}=1^{-+}$ are inconsistent with a simple 
quark-antiquark bound state.
Two such exotics, denoted $\pi_1$, have been observed through
$\pi^- p \rightarrow \pi_1 p$ at BNL \cite{exoticb}:
with masses 1400 MeV (in decays to $\eta \pi$) 
and 1600 MeV (in decays to $\eta' \pi$ and $\rho \pi$).
The $\pi_1 (1400)$ state has also been observed in ${\bar p} N$
processes by the Crystal Barrel Collaboration at CERN \cite{exoticc}. 
Exotic quantum numbers can be generated through a
``valence'' gluonic component 
-- for example through coupling 
to the operator $[ {\bar q} \gamma_{\mu} q G^{\mu \nu}$ ].
However, the observed states are considerably lighter than 
the predictions 
(about 1800-1900 MeV) of quenched lattice QCD 
\cite{lattice} and  QCD inspired models \cite{models} 
for the lowest mass $q {\bar q} g$ state with $J^{PC}=1^{-+}$.
These results suggest that, 
perhaps, the ``exotic'' states observed by the experimentalists 
might involve significant meson-meson bound state contributions.  
Furthermore,
the decays of the light mass exotics to $\eta$ or $\eta'$ mesons 
plus a pion may hint at a possible connection to $U_A(1)$ dynamics.
This idea has recently been investigated \cite{bassem} in a model of 
final state interaction in $\eta \pi$ and $\eta' \pi$ scattering 
using coupled channels and the Bethe-Salpeter equation following 
the approach in \cite{valencia}.
In this calculation the meson-meson scattering potentials were derived 
from the $U_A(1)$ extended chiral Lagrangian working to $O(p^2)$ in the 
meson momenta.
Fourth order terms in the meson fields induced 
the OZI violating interaction
$
\lambda \ Q^2 \ {\rm Tr} \ \partial_{\mu} U \partial^{\mu} U^{\dagger}
$
\cite{veccb}
are found to play a key role.
Here %$Q$ is the topological charge density and 
$U$ is the unitary 
meson matrix associated with spontaneously broken chiral symmetry.
A simple estimate for $\lambda$ can be deduced from the decay 
$\eta' \rightarrow \eta \pi \pi$ 
yielding two possible solutions with different signs.
Especially interesting is the negative sign solution.
When substituted into the Bethe-Salpeter equation
this solution yielded a dynamically generated p-wave 
resonance with exotic quantum numbers $J^{PC}=1^{-+}$, 
mass $\sim 1400$ MeV and width $\sim 300$ MeV -- close to the observed 
exotics.
%(The width of the $\pi_1(1400)$ state measured in decays to $\eta \pi$ 
% is $385 \pm 40$MeV; the width of the $\pi_1(1600)$ measured in decays
% to $\eta' \pi$ is $340 \pm 64$MeV.)
The topological charge density mediates the coupling of the 
dynamically generated light-mass exotic to the $\eta \pi$ and 
$\eta' \pi$ channels in this model \cite{bassem}.

\end{document}